# Definition of Power Converters

*F. Bordry and D. Aguglia*
CERN, Geneva, Switzerland

**Abstract**
The paper is intended to introduce power conversion principles and to define common terms in the domain. The concepts of sources and switches are defined and classified. From the basic laws of source interconnections, a generic method of power converter synthesis is presented. Some examples illustrate this systematic method. Finally, the commutation cell and soft commutation are introduced and discussed.

**Keywords**
Power converter; power electronics; semiconductor switches; electrical sources; design rules; topologies.

## 1 Introduction

The task of a power converter is to process and control the flow of electrical energy by supplying voltages and currents in a form that is optimally suited for user loads.

Energy conversions were initially achieved using electromechanical converters (which were mainly rotating machines). Today, with the development and the massive production of power semiconductors, static power converters are used in numerous application domains and especially in particle accelerators. Their weight and volume are smaller and their static and dynamic performance are better.

A static converter is composed of a set of electrical components building a meshed network that acts as a linking, adapting, or transforming stage between two sources, generally between a generator and a load (Fig. 1).

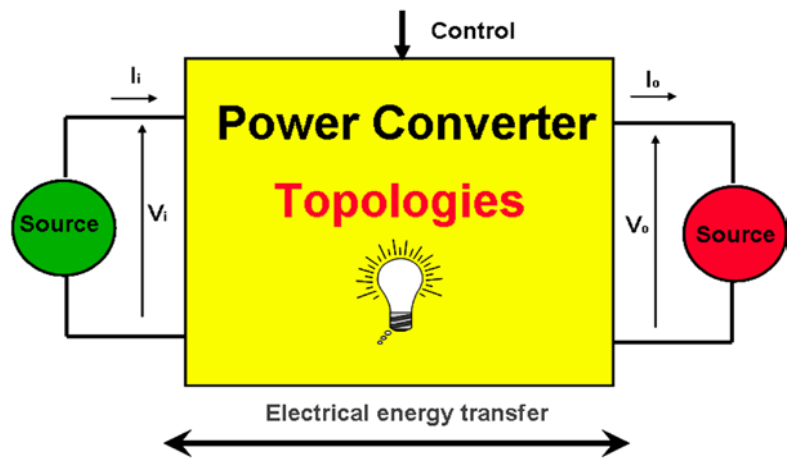

**Fig. 1:** Definition of a power converter

An ideal static converter allows control of the power flow between the two sources with 100% efficiency. A large part of power converter design is the optimization of its efficiency. But as a first approach and to define basic topologies, it is interesting to take the hypothesis that no losses occur

through a power converter's conversion process. With this hypothesis, the basic elements are of two types:

- non-linear elements that are, most of the time, electronic switches: semiconductors used in commutation mode [1];
- linear reactive elements: capacitors, inductances (and mutual inductances or transformers). These reactive components are used for intermediate energy storage but also perform voltage and current filtering. They generally represent an important part of the size, weight and cost of the equipment [2, 3].

The objective of this introductory paper is to recall and give a precise definition of basic concepts essential for the design or understanding of power converter topologies [4]. First, a very simple example is presented to illustrate the basic principle of modern power electronics converters (switching power converter). Then the sources and the switches are defined, followed by the fundamental connection rules between these basic elements. From there, converter topologies are deduced. Some examples of topology synthesis are then given. Finally, the concept of hard and soft commutation is introduced.

## 2   The very basic principles of modern power electronics conversion

As a first basic example to illustrate the evolution toward modern power electronics, let us consider a DC to DC converter that aims to deliver 100 V to a resistive load of 10 Ω. The input voltage source delivers a constant 325 V. Until the 1960s, and still in use in some special applications, a typical utilization of transistors consisted of operating them in their linear, or active, region. The basic 'old' topology for this case is illustrated in Fig. 2. This consists of operating the transistor such that a 225 V voltage drop across it is ensured. Simplifying, one can say that in this case the transistor is used as a controllable resistance.

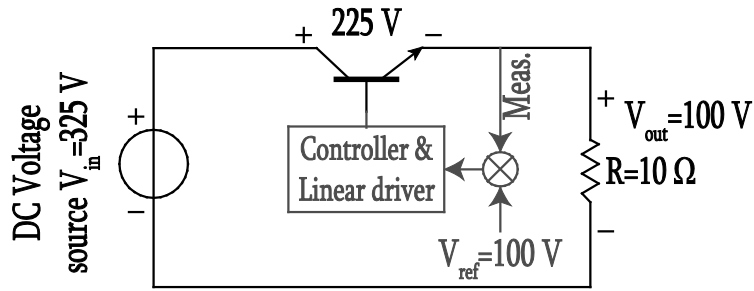

**Fig. 2:** DC to DC converter using the transistor's linear, or active, region

Analysing this circuit at this operating point, one can derive the input and the delivered output powers $P_{in}$ and $P_{out}$, the losses in the transistor $P_T$, and the converter efficiency $\eta$, as presented in Eq. (1),

$$P_{in} = 325 \text{ V} \times 10 \text{ A} = 3.25 \text{ kW}, \tag{1}$$

$$P_{out} = 100 \text{ V} \times 10 \text{ A} = 1 \text{ kW}, \tag{2}$$

$$P_T = P_{in} - P_{out} = 2.25 \text{ kW}, \tag{3}$$

$$\eta = \frac{P_{out}}{P_{in}} = 0.3 \equiv 30\%. \tag{4}$$

Notice that for this particular operating point the efficiency is extremely low. In the general case, the efficiency can be expressed as in Eq. (5), where the efficiency drastically decreases as $V_{out}$ decreases,

$$\eta = \frac{P_{out}}{P_{in}} = \frac{V_{out}^2/R}{(V_{in} \cdot V_{out})/R} = \frac{V_{out}}{V_{in}}. \tag{5}$$

This conversion system produces high losses and needs an over-dimensioned transistor able to dissipate the losses. Furthermore, the overall dimensions of this power converter shall be large enough for being able of evacuating the high losses for a given maximum temperature. This conversion method is still used in some special applications where high precision or high dynamics is required [5].

The basic principle of modern power electronics lies in the utilization of switches in their ON and OFF states only, virtually producing no, or very low, losses (losses in switches are described in Section 4.2, Dynamic characteristics). The basic principle is illustrated in Fig. 3. Between the voltage source $V_{in}$ and the load $R$, the converter is now composed of a 'special' two-position switch and a low-pass filter. When the switch is in position 1, the input voltage $V_{in}$ appears at the low-pass filter input $V_s$. When the switch is in position 2, zero volts are applied to $V_s$. Given a pattern in time of the switch positions, one derives the pattern of the voltage $V_s$ (chopped or switched voltage). In most applications, the voltage $V_s$, with a fluctuation of 100% (from 0 V to $V_{in}$), cannot be directly applied to the load, therefore a filtering action must be undertaken in order to apply the voltage $V_s$ average values to the load, which is derived in Eq. (6),

$$V_{\text{out}} = \frac{1}{T_s}\int_0^{T_s} v_{in}(t)\mathrm{d}t = DV_{\text{in}}, \qquad (6)$$

where $T_s$ is the switching period (which defines the switching frequency $f_s$), and $D$ is the duty cycle, or duty ratio, which defines the relative time when the switch is in position 1 with respect to the switching period $T_s$ (illustrative definitions are given in Fig. 3). In this case, the regulation of the output voltage is performed by acting on the duty cycle $D$.

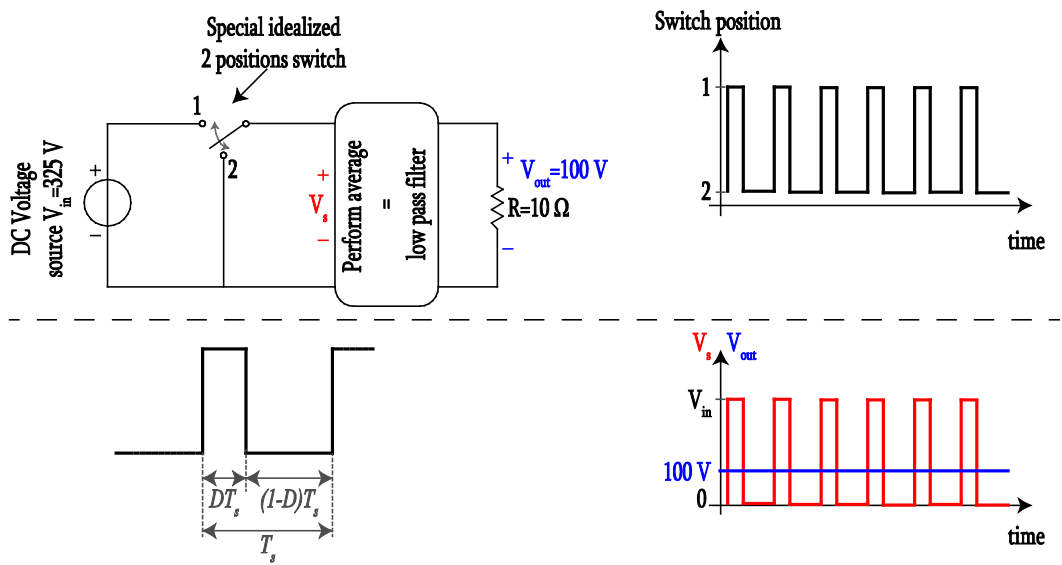

**Fig. 3:** Idealized switch-mode DC-to-DC converter principle

Thanks to the low losses, modern switch-mode power converters are more efficient and compact. However several new aspects have to be considered during the design phase. A filtering process is required, leading to the introduction of the so-called output voltage and current ripples on the load and voltage, or current, bandwidths. A non-ideal filter is always letting some harmonics pass through the load. Furthermore, the dynamic characteristics of the power converter (rate of change of current and/or voltage) is limited by the filter. This trade-off between ripple and bandwidth is a key aspect in the design and specification process of modern switch-mode power converters.

## 3 Sources

As mentioned in the introduction, a power converter processes the flow of energy between two sources. To synthesise a power converter topology, the first step is to characterize these sources. We will see later that the converter structure can be directly deduced as soon as the sources are defined: voltage or current sources and their reversibilities.

In the energy conversion process, a source is usually a generator (often called an input source) or a load (often called an output source). However, in the case of a change of direction of the energy flow, i.e. a change in the sign of the power, the sources (generators and loads) can exchange their functions (i.e. restoration of energy from a magnet back to the grid).

## 3.1 Nature of sources

### 3.1.1 Definitions

Two types of sources could be defined: voltage and current sources. As mentioned, any of these sources could be a generator or a receiver (load).

A source is called a *voltage* source if it is able to impose a voltage independently of the current flowing through it. This implies that the series impedance of the source is zero (or negligible in comparison with the load impedance)

A source is called a *current* source if it is able to impose a current independently of the voltage at its terminals. This implies that the series impedance of the source is infinite (or very large in comparison with the load impedance).

These definitions correspond to permanent properties. The principle of operation of a converter is based on the switch-mode action of its switches. Commutation of the switches generates very fast current and/or voltage transients so that the transient behaviour of the sources is of fundamental importance for the converter design. The transient behaviour of a source is characterized by its ability or inability to withstand steps in the voltage across its terminal or in the current flowing through it, these steps being generated by the external circuit. Then new definitions could be stated as below.

- A source is a *voltage source* if the voltage across its terminals cannot undergo a discontinuity due to the external circuit variation. The most representative example is a capacitor, since an instantaneous change of voltage would correspond to an instantaneous change of its charge that would require an infinite current (Fig. 4(b)).
- A source is a *current source* if the current flowing through it cannot undergo a discontinuity due to the external circuit variation. The most representative example is an inductor, since an instantaneous change in current would correspond to an instantaneous change in its flux that would require an infinite voltage (Fig. 4(a)).

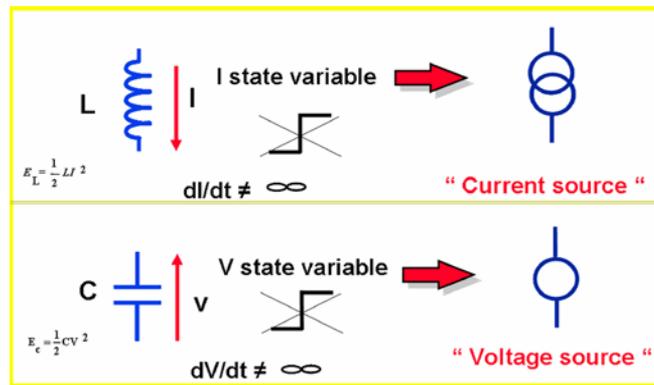

**Fig. 4:** Inductance and capacitor vs. current and voltage source

It should be noted that a square wave voltage generator is indeed a voltage source as defined above since the voltage steps are not caused by the external circuit. A square wave current generator is indeed a current source as defined above since the current steps are not caused by the external circuit.

With these definitions, it is interesting to define the notion of instantaneous impedance of a source as the limit of the source impedance when the Laplace operator tends towards infinity. Theoretically this instantaneous impedance can be zero, finite, or infinite.

A source is referred to as a voltage source when its instantaneous impedance is zero, while a source is called a current source if its instantaneous impedance is infinite.

For example:

Capacitor: $Z(s) = 1/(C.s)$, $\lim_{s \to \infty} Z(s) = 0$, this leads to a voltage source;

Inductance: $Z(s) = L.s$, $\lim_{s \to 0} Z(s) = \infty$, this leads to a current source.

### 3.1.2 Source reversibility

The determination of the source reversibilities is fundamental. We will see that the reversibility analysis allows the deduction of the static characteristics of the switches.

The voltage (or the current) that characterizes a source is termed DC if it is unidirectional. As a first approximation, it can be taken as being constant. The voltage (or the current) is termed AC if it is periodic and has an average value equal to zero. As a first approximation, it can be taken as sinusoidal.

A source is voltage-reversible if the voltage across its terminals can change sign. In the same way, a source is current-reversible if the current flowing through it can reverse.

In summary, the input/output of a converter can be characterized as voltage or current sources (generator or loads), either DC or AC, current-reversible and/or voltage-reversible. In total, there are only eight possibilities, shown in Fig. 5.

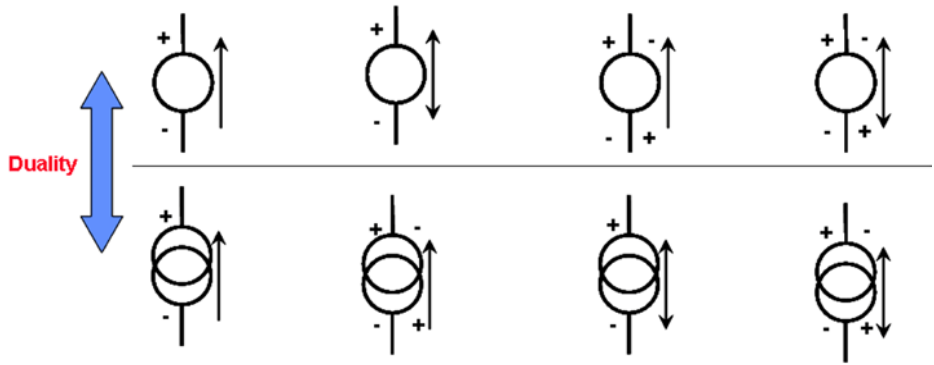

**Fig. 5:** Voltage and current sources with their reversibilities

### 3.1.3 Source nature modification

Connection of a series inductance with an appropriate value to a voltage source (i.e. a dipole with zero instantaneous impedance) turns the voltage source into a current source. In the same way, connecting a parallel capacitor of appropriate value to a current source (a dipole with infinite instantaneous impedance) turns the current source into a voltage source (Fig. 6).

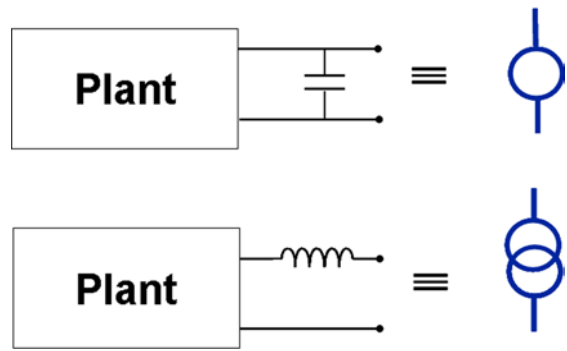

**Fig. 6:** Source nature confirmation or modification

These inductive or capacitive elements connected in parallel or in series with the source are elements that can temporarily store energy. Consequently, if an inductance connected to a voltage source turns it into a current source it is important to determine the current reversibility of this current source.

In practice the identification of a real generator or of a real load as a voltage or current source is not obvious. That is the reason why the nature of the source is often reinforced by the addition of a parallel capacitor in the case of voltage sources and by the addition of a series inductor in the case of current sources.

Obviously, the current source obtained by connecting an inductance in series with a voltage source keeps the same current reversibility as this voltage source. The inductance acts as a buffer absorbing the voltage differences. Consequently the current source obtained by connecting a series inductance to a voltage source is reversible in voltage. When the voltage source itself is reversible in voltage there is no particular problem. But, if the voltage source is not reversible in voltage, the current source obtained by connecting a series inductance to the voltage source is only instantaneously reversible with respect to voltage.

The former result can easily be transposed to the voltage source obtained by the parallel connection of a capacitor to a current source. The voltage source obtained keeps the same voltage reversibility as the current source and is reversible in current. However, this reversibility is only instantaneous if the current source is not reversible in current.

### 3.1.4 *Example*

A set of ideal batteries behaves as a load during charging and as a generator during discharging; such a source is called a DC voltage source, being current reversible but not voltage reversible. Nevertheless, because of the inductance of the connecting cables, this battery can sometimes be taken as a current source that is instantaneously voltage reversible and permanently current reversible. If a capacitor bank is added at the terminals of the cables, it again becomes a voltage source (Fig. 7).

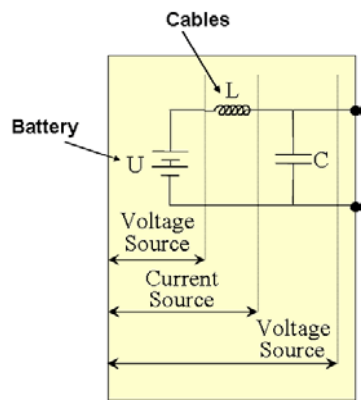

**Fig. 7:** Modification to voltage or current source

## 4 Switch characteristics

Static converters are electrical networks mainly composed of semiconductor devices operating in switch mode (as switches). Through proper sequential operation of these components, they allow an energy transfer between two sources with different electrical properties.

The losses in the switches should be minimized in order to maximize the efficiency of the converter. Switches must have a voltage drop (or an ON resistance) as low as possible in the ON state, and a negligible leakage current (or an OFF resistance) in the OFF state. These two states are defined as static states.

The change from one state to the other state (switch commutation) implies transient behaviour of the switch. These behaviours are complex because they depend on the control of the switch (through a gate control) and on the conditions imposed by the external circuit.

## 4.1 Static characteristics

In the static domain a switch has the same behaviour as a non-linear resistance: very low in the ON state and very high in the OFF state.

Taken as a dipole with the load sign convention (Fig. 8) the static characteristic $I_k(V_k)$, which represents the operating points of a switch, is made up of two branches totally located in quadrants 1 and 3 such that $(V_k \times I_k) > 0$. One of these branches is very close to the $I_k$ axis (ON state) and the other is very close to the $V_k$ axis (OFF state). Each of these branches can be located in one or two quadrants. In the case of an ideal switch, the static characteristics are the half-axis to which they are close.

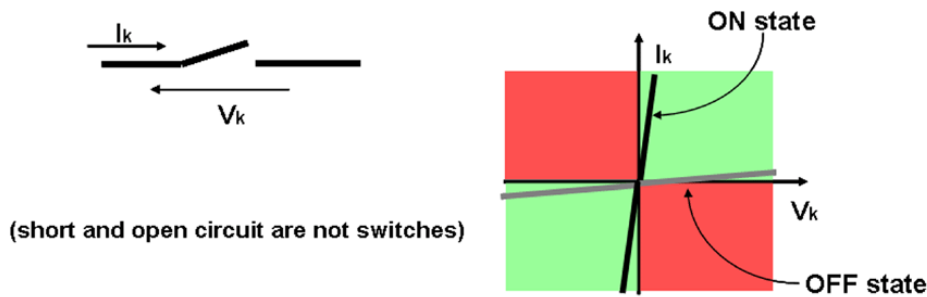

**Fig. 8:** Static characteristics of a switch

In this representation, except for the obvious cases of a short circuit and of an open circuit that correspond respectively to a switch always ON and to a switch always OFF, any switch that really behaves as a switch (commutation: ON <=> OFF) has a static characteristic consisting of at least two orthogonal half-axes (or segments).

The static characteristic, an intrinsic feature of a switch, reduces to a certain number of segments in the $I_k(V_k)$ plane.

– Two-segment characteristics: the switch is unidirectional in current and in voltage. Two two-segment characteristics can be distinguished: in the first case, current $I_k$ and voltage $V_k$ have the same signs; in the second case, current $I_k$ and voltage $V_k$ have opposite signs. The switches having such characteristics are respectively called T and D switches (Fig. 9).

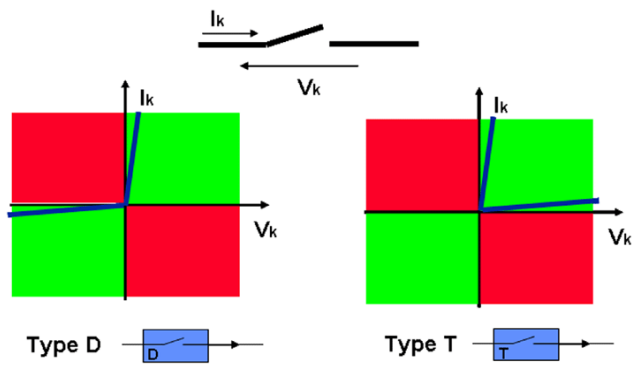

**Fig. 9:** Static characteristics of a two-segment switch

– Three-segment characteristics: the switch is bidirectional either in current or in voltage while the other is unidirectional. Therefore, there are two types of three-segment static characteristics (Fig. 10). It should be noted that these two types of switches could be synthesized with the association in parallel or in series of two-segment switches (T and D).

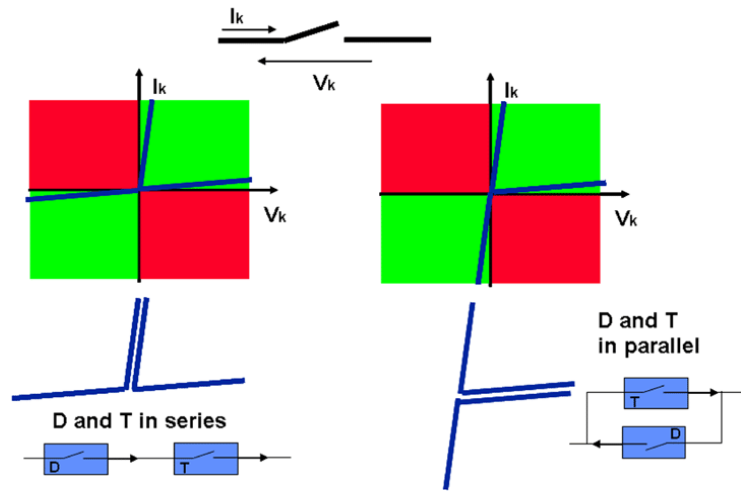

**Fig. 10:** Static characteristics of a three-segment switch

– Four-segment characteristics: the switch is bidirectional in voltage and in current. There is only one such type of static characteristic (Fig. 11). A four-segment characteristic could be obtained by series or parallel connection of switches with three-segment characteristics.

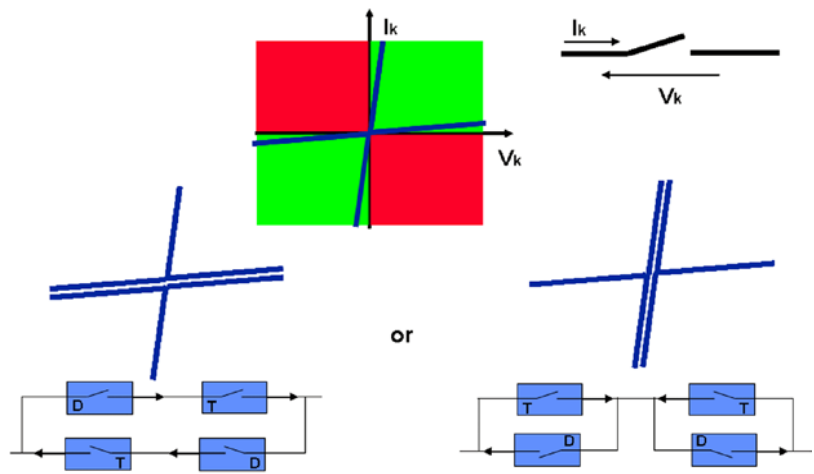

**Fig. 11:** Static characteristics of a four-segment switch

### 4.2 Dynamic characteristics

The dynamic characteristic is the trajectory described by the point of operation of the switch during its commutation, to go from one half-axis to the perpendicular half-axis. A switch being either ON or OFF, there are two commutation dynamic characteristics corresponding to the turn-ON and the turn-OFF, which will be grouped under the global term dynamic characteristics.

Unlike the static characteristic, the dynamic characteristic is not an intrinsic property of the switch but also depends on the constraints imposed by the external circuit. Neglecting second-order phenomena, and taking into account the dissipative nature of the switch, the dynamic characteristic can only be located in those quadrants where $V_k \times I_k > 0$ (generator quadrants). For the two commutations (turn-ON and turn-OFF), two modes are possible: controlled commutation and spontaneous commutation.

*4.2.1 Controlled commutation*

The switch has, in addition to its two main terminals, a control terminal on which it is possible to act in order to provoke a quasi-instantaneous change of state (in the case of a T switch). The internal resistance of this switch can change from a very low value to a very high value at turn-OFF (and inversely at turn-ON). These changes are independent of the evolution of the electrical quantities imposed on the switch by the external circuit.

It should be noted that, in a controlled commutation, the switch imposes its state on the external circuit. Under such circumstances, the element can undergo severe stresses that depend on its dynamic characteristic. If the switching time is long and the operating frequency is high, the commutation losses can be important.

*4.2.2 Spontaneous commutation*

The spontaneous commutations correspond to turn-OFF when the current flowing through the switch arrives at zero and to turn-ON when the voltage applied across its terminals reaches zero. Spontaneous commutation is the commutation of a simple PN junction (D switch). It is only dependent on the evolution of the electrical variables in the external circuit. Spontaneous commutations could be achieved with any controlled semiconductor if the gate control is synchronized with the electrical quantities of the external circuit. Spontaneous commutation is achieved with minimal losses since the operating point moves along the axes.

It is important to point out that controlled commutation can only happen in the first or third quadrants while spontaneous commutation can only happen with a change of quadrant (Fig. 12).

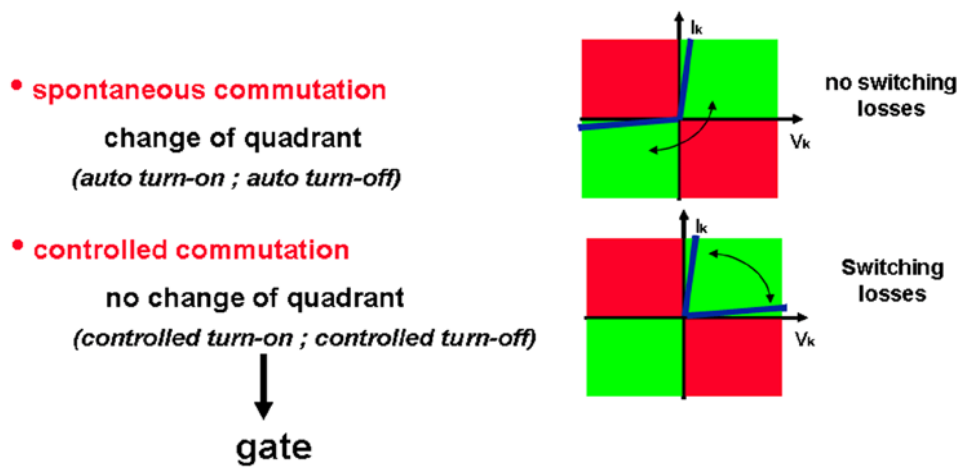

**Fig. 12:** Spontaneous and controlled commutations

## 4.3 Classification of switches

Finally, switches used in power converter can be classified by their static characteristics (two, three or four segments) and by the type of commutation (controlled or spontaneous) at turn-ON and at turn-OFF.

*4.3.1 Two-segment switches*

Except the open circuit and the short circuit, two switches with two-segment characteristics can be distinguished (Fig. 13).

– The first of these switches has the static characteristics of switch D, and its turn-ON and turn-OFF commutations are spontaneous. This switch is typically a diode.

– The second of these switches has the static characteristics of switch T and its turn-ON and turn-OFF commutations are controlled. Examples are the power semiconductors: metal-oxide semiconductior field-effect transistors (MOSFET), insulated-gate bipolar transistor (IGBT), gate tunr-off thyristor (GTO), integrated-gate commutated thyristor (IGCT), etc.

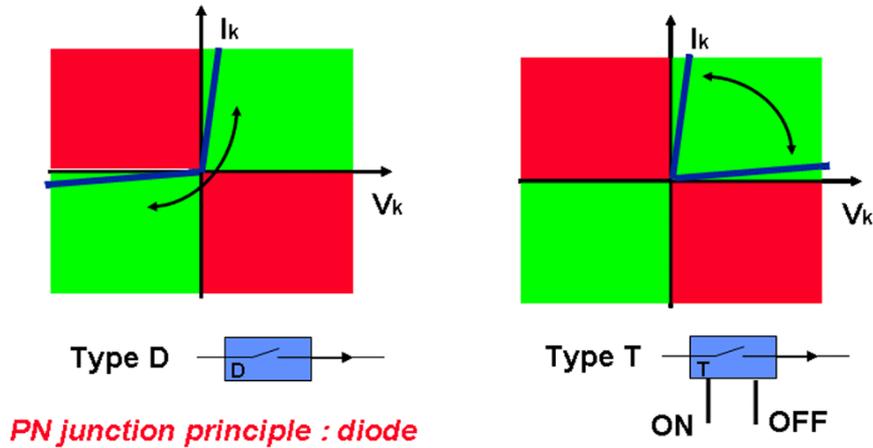

**Fig.13:** Dynamic characteristics of two-segment switches

This switch will be symbolized by separating the turn-ON and turn-OFF control gate as shown in Fig. 13.

A switch with two-segment characteristics similar to the T type (both segments in the same quadrant), must have controlled turn-ON and turn-OFF commutations. If it would have only one controlled commutation, it would be necessary to put in series or in parallel a switch D (a diode) to get the spontaneous commutation. In this case, it is no longer a two-segment switch, but a three-segment one. Therefore, only two two-segment switches can be used directly.

*4.3.2 Three-segment switches*

These switches can be divided into two groups depending whether they are:

– unidirectional in current and bidirectional in voltage (Fig. 14);
– bidirectional in current and unidirectional in voltage (Fig. 15).

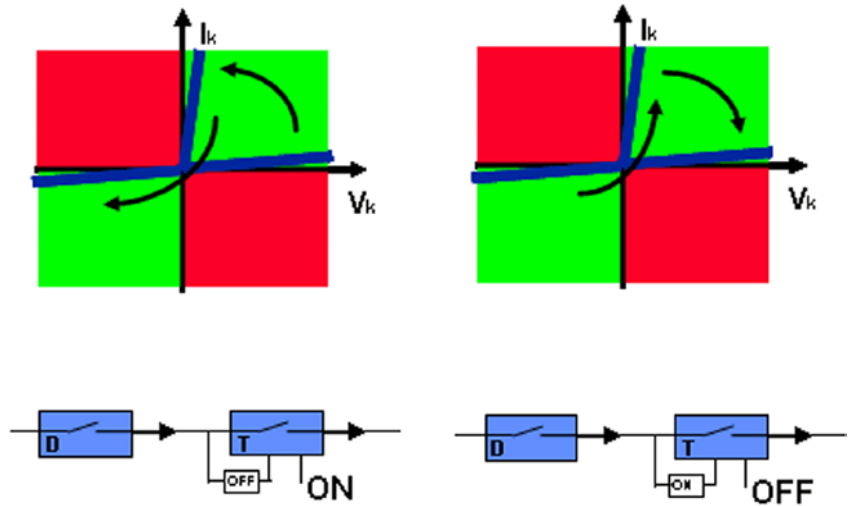

**Fig, 14:** Dynamic characteristics of three-segment switches: bidirectional in voltage

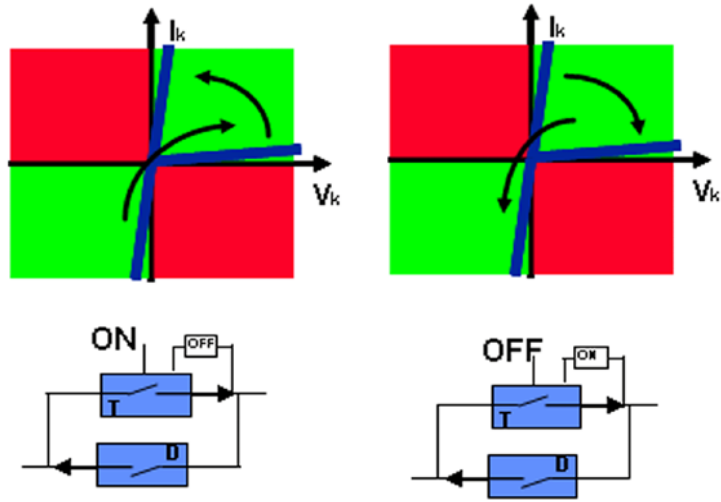

**Fig. 15:** Dynamic characteristics of three-segment switches: bidirectional in current

Except for the thyristor, all these switches are synthesized switches, realized by connecting a diode in parallel or in series with a two-segment switch. A special driver is needed to obtain spontaneous commutation. The 'dual-thyristor' (unidirectional in voltage, bidirectional in current, controlled turn-OFF and spontaneous turn-ON) is a good example of a useful three-segment switch [6, 7].

In each of these two groups, all switches have the same static characteristics but differ with their commutation mechanisms. It is important to remark that if a three-segment switch has both commutations controlled (turn-ON and turn-OFF) or both spontaneous, it would never use the three segments of its static characteristic. Therefore, a three-segment switch must necessarily have one controlled commutation and one spontaneous commutation.

The cycle of operation, which represents the locus described by the point of operation of these switches, is then fully determined. They can only be used in converter topologies that impose a single cycle on the switches during operation.

*4.3.3   Four-segment switches*

All four-segment switches have the same static characteristic. They differ only by their commutation modes that can, a priori, differ in quadrants 1 and 3. So, six four-segment switches can be distinguished.

These switches are used mainly in direct frequency changers and in matrix converters; in practice they are made up of two three-segment switches connected in series or parallel.

## 5   Interconnection of sources: Commutation rules

To control the power flow between two sources, the operation principle of a static power converter is based on the control of switches (turn-ON and turn-OFF) with determined cycles creating periodic modifications of the interconnection between these two sources.

The source interconnection laws can be expressed in a very simple way:

– a voltage source should never be short-circuited but it can be open-circuited;
– a current source should never be open-circuited but it can be short-circuited.

From these two general laws, it can be deduced that switches cannot establish a direct connection between two voltage sources or between two current sources.

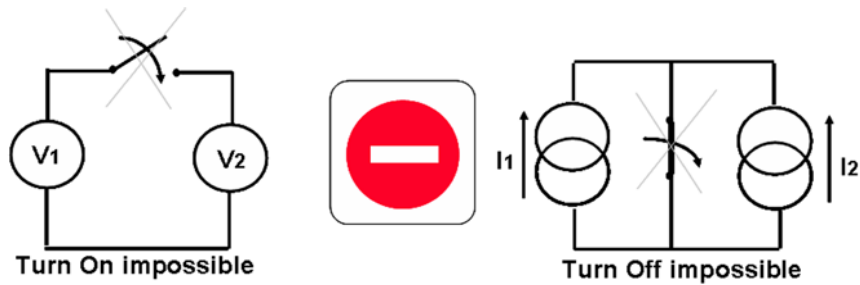

**Fig. 16:** Basic interdictions of source interconnection

In the case of two voltage sources, the switch turn-ON can only happen when the two sources have the same values, that is to say at the zero crossing of the voltage across the switch. The turn-ON must then be spontaneous (since it depends on the external circuit) and turn-OFF can be controlled at any time. In the case of two current sources, the switch turn-OFF can only happen when the two current sources reach the same value, that is to say when the current in the switch reaches zero. In this case, the turn-OFF is spontaneous and turn-ON can be controlled at any time.

As it is done in day-to-day practise and following the previous arguments, it is obvious that it is possible to put capacitors in parallel and inductances in series but these elements should be connected with zero voltage or zero current, respectively.

From the laws state above, it is easy to deduce that the commutation of switches must not interconnect two sources of the same nature.

## 6  Structure of power converters

A power converter can be designed with different topologies and with one or several intermediate conversion stages. When this conversion is achieved without any intermediate stage temporarily storing some energy, the conversion is called direct conversion and it is achieved by a direct converter. On the other hand, when this conversion makes use of one or more stages able to store energy temporarily, the conversion is termed indirect and it is achieved by an indirect converter.

The interdiction to connect two sources of the same nature leads to the consideration of two classes of basic conversion topologies:

– direct link topology: when the two sources have different natures;
– indirect link topology: when the two sources have the same nature.

### 6.1  Direct link topology converters

A direct converter is an electrical network composed only of switches and is unable to store energy. In such a converter, the energy is directly transferred from the input to the output (with the hypothesis that the losses can be neglected); the input power is at any time equal to the output power.

Taking into account the interconnection rules recalled above, the different possible connections between a voltage source and a current source are shown in Fig. 17. To get all these connections, the simplest structure is the four-switches bridge (Fig. 18):

– with K1 and K3 closed, the connection shown in Fig. 17(a) is obtained;
– with K2 and K4 closed, the connection shown in Fig. 17(b) is obtained;
– with K1 and K2 closed (or with K3 and K4 closed), the connection shown in Fig. 17(c) is obtained.

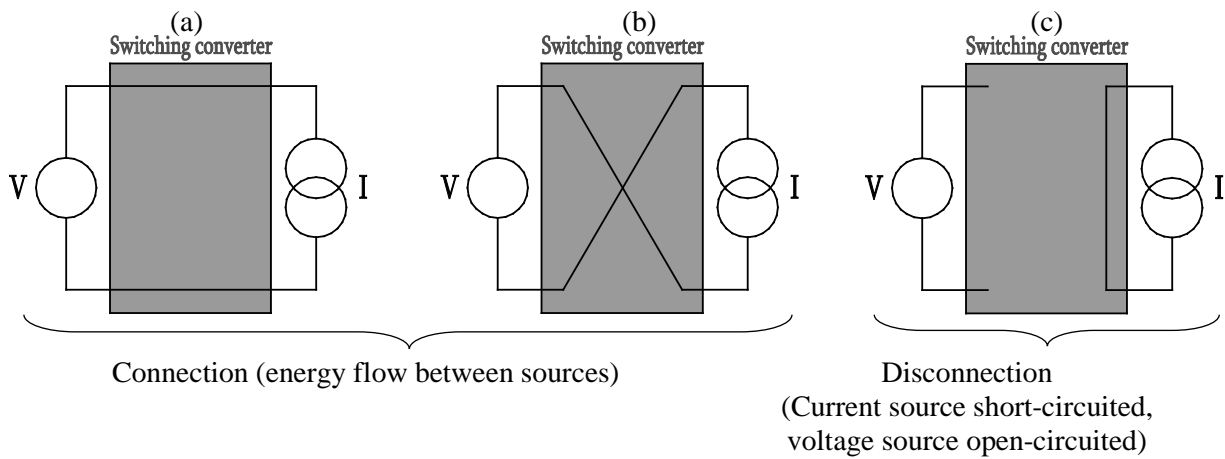

**Fig. 17:** Interconnection possibilities between a voltage and a current source. (a) – direct connection; (b) – inverse connection; (c) - disconnection with current source in short-circuit.

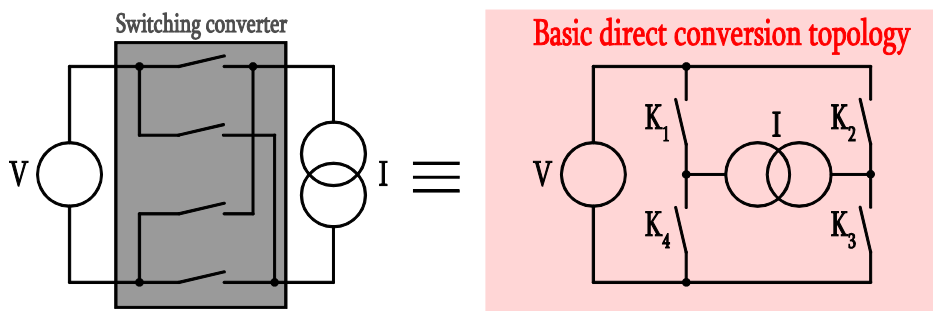

**Fig. 18:** Basic configuration of voltage–current direct converter

When some of these connections are not necessary, it is possible to simplify the bridge structure into structures using fewer switches (i.e. a buck converter).

Energy conversion between an input current source and an output voltage source is the same problem. The basic configuration is the same but it is more usual to represent the input source on the left and the output source on the right (Fig. 19).

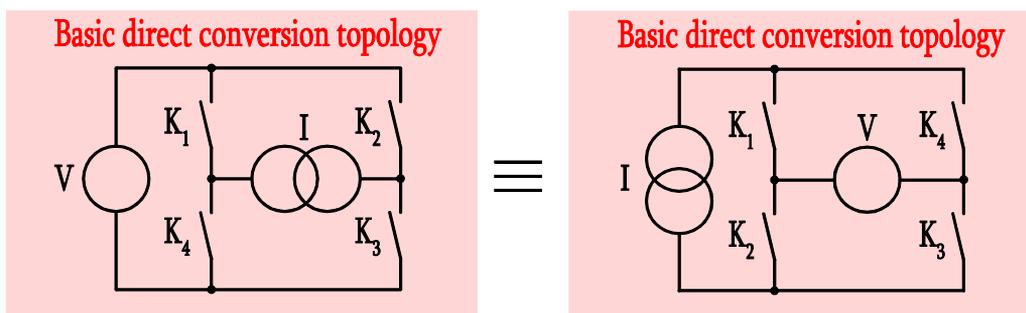

**Fig. 19:** Basic configuration of current–voltage direct converter

## 6.2 Indirect converters

If both sources have the same nature, it is not possible to interconnect them directly with switches. It is necessary to add components to generate an intermediate buffer stage of a different type without active energy consumption (capacitor or inductor). This buffer stage is a voltage source (capacitor) if the energy transfer is between two current sources, and it is a current source (inductance) if the energy transfer is between two voltage sources.

### 6.2.1 Modification of the nature of the input or output source

In the case of voltage–voltage conversion, one solution could be to add an inductance in series with the input voltage source or with the output voltage source. With this change of the source nature, it is possible to use a direct converter: a current–voltage or voltage–current converter according to where the inductance was added (Fig. 20).

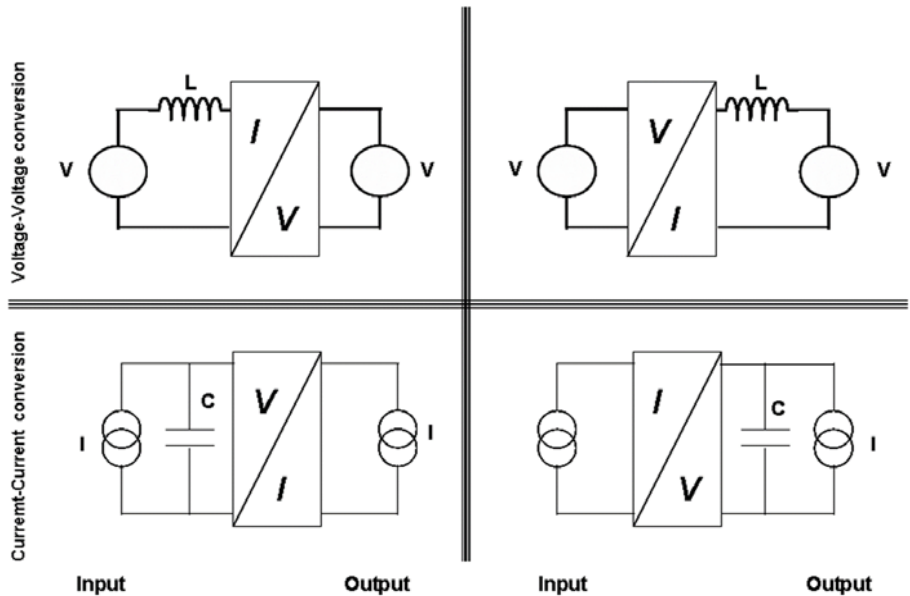

**Fig. 20:** Indirect converters: modification of the nature of the input or output source. Above one of the voltage sources need to be modified into a current source. Below one of the current sources need to be modified into a voltage source.

The case of current–current conversion is similar to the previous case. Therefore, a capacitor should be added in parallel or in series with the input current source or with the output current source.

### 6.2.2 Use of two direct converters

If it is not possible or too costly to modify the nature of one source, two direct converters can be connected with an intermediate buffer stage: an inductance for a voltage–voltage conversion and a capacitor for a current–current conversion (Fig. 21).

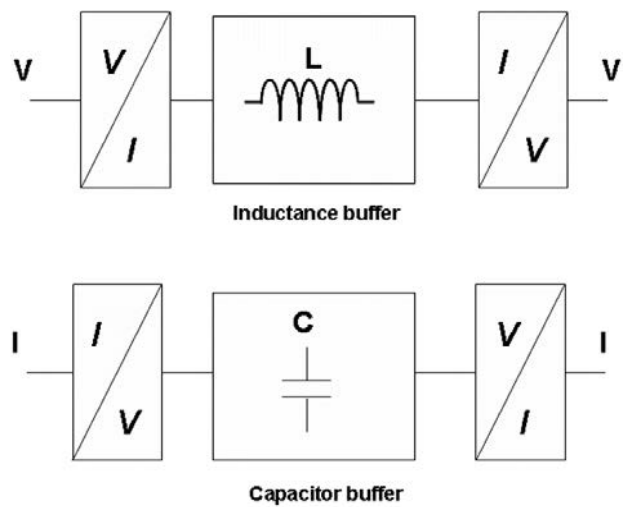

**Fig. 21:** Indirect converters: intermediate buffer stage, inductive above and capacitive below, between two direct converters.

### 6.2.3 Voltage–voltage indirect converters

In indirect converters, the two voltage sources are never connected directly (see the basic laws described above). Two sequences are therefore necessary. During the first sequence, the energy is transferred from the input voltage source to the inductance (voltage to current conversion). During the second sequence, the inductance gives back the energy to the output voltage source (current to voltage conversion; two directions are possible) (Fig. 22). An extra switch is necessary to obtain these sequences.

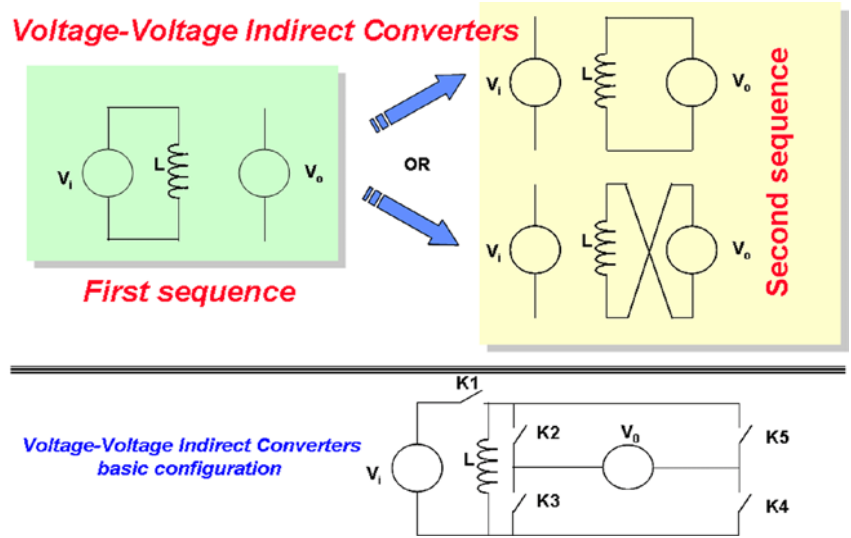

**Fig. 22:** Voltage–voltage indirect converters. Above the three possible connections between first and second sequences. Below the corresponding topology.

### 6.2.4 Current–current indirect converters

In the indirect converters, the two current sources are never connected directly. Then, two sequences are necessary. During the first sequence, the energy is transferred from the input current source to the capacitor (current to voltage conversion). During the second sequence, the capacitor gives back the energy to the output current source (voltage to current conversion; two directions are possible) (Fig. 23).

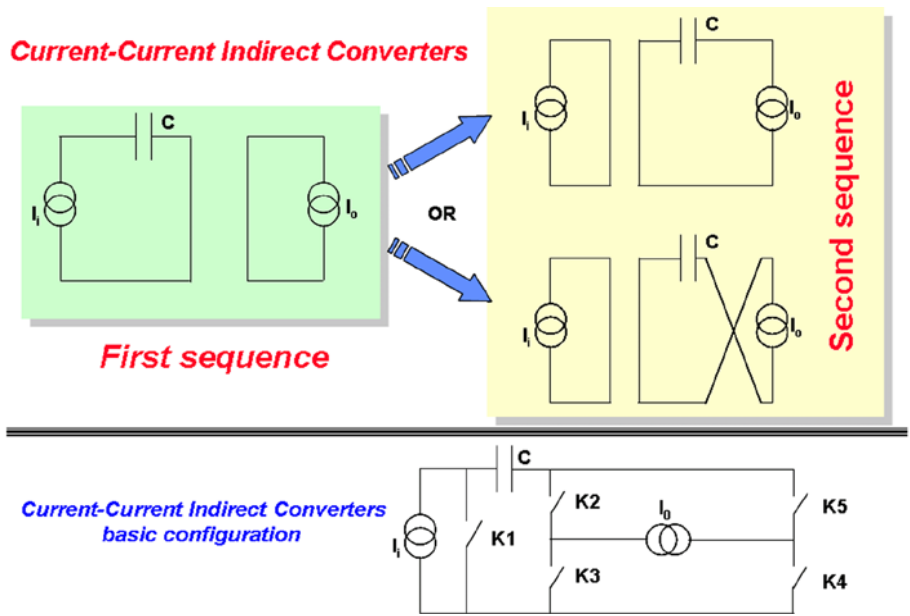

**Fig. 23:** Current–current indirect converters. Above the three possible connections between first and second sequences. Below the corresponding topology.

## 6.3 Conclusions

Figure 24 represents the three basic configurations of all single-phase converters.

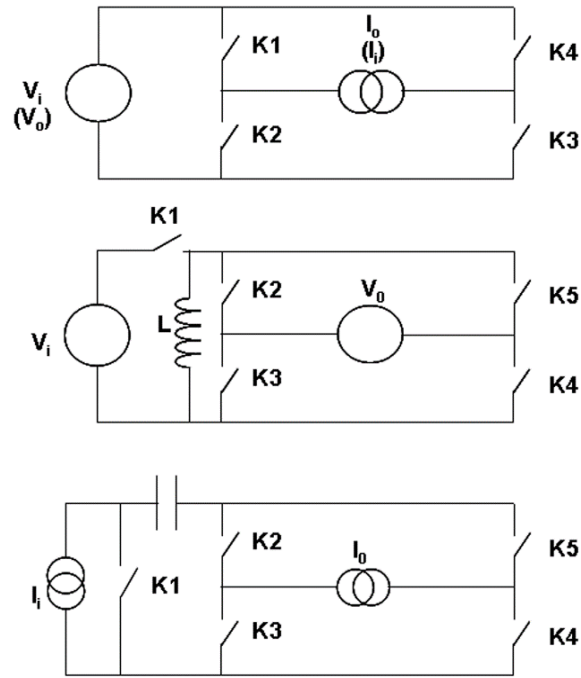

**Fig. 24:** Three basic structures of single-phase power converters. Voltage to current (and vice-versa) direct conversion (above), voltage to voltage indirect conversion (in the middle), and current to current indirect conversion (below).

From these basic configurations, it is possible, according to the nature of the sources, to associate them or to add other components. For example, in the case where one of the sources is AC, it is possible to insert a transformer for adaptation or galvanic insulation (Fig. 25).

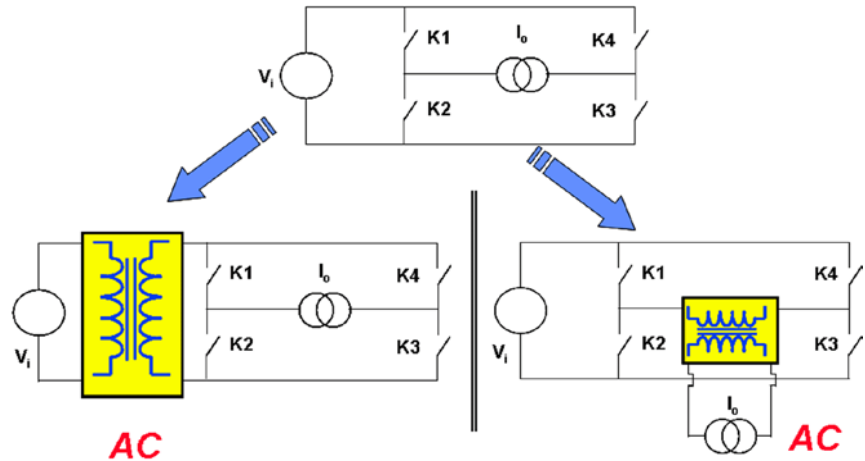

**Fig. 25:** Insertion of a transformer in a direct topology

It is also possible to associate several basic topologies. One application that is more frequently used is to have an intermediate stage operating at a high frequency to reduce the size of the transformers and magnetic elements and to get higher performance at the output (bandwidth, ripple, etc.) (Fig. 26). Details of magnetic component design are given in Ref. [3].

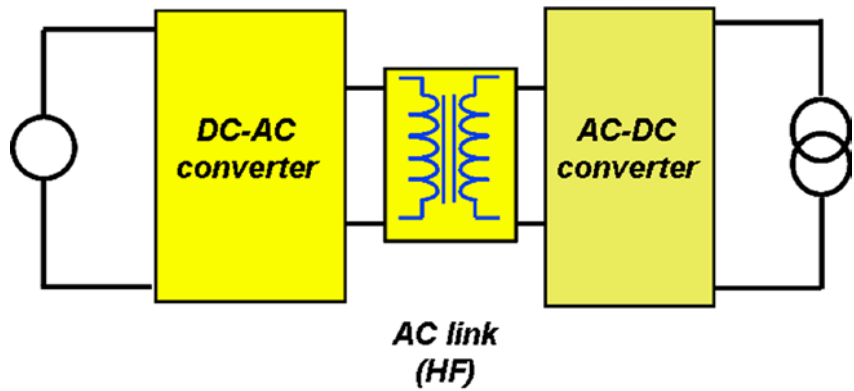

**Fig. 26:** Association of elementary structures

Figure 27 illustrates the case of resonant converters [8, 9]. It should be noted that the interconnection of the various intermediate sources must respect the interconnection laws. It is especially important for the choice of the output filter.

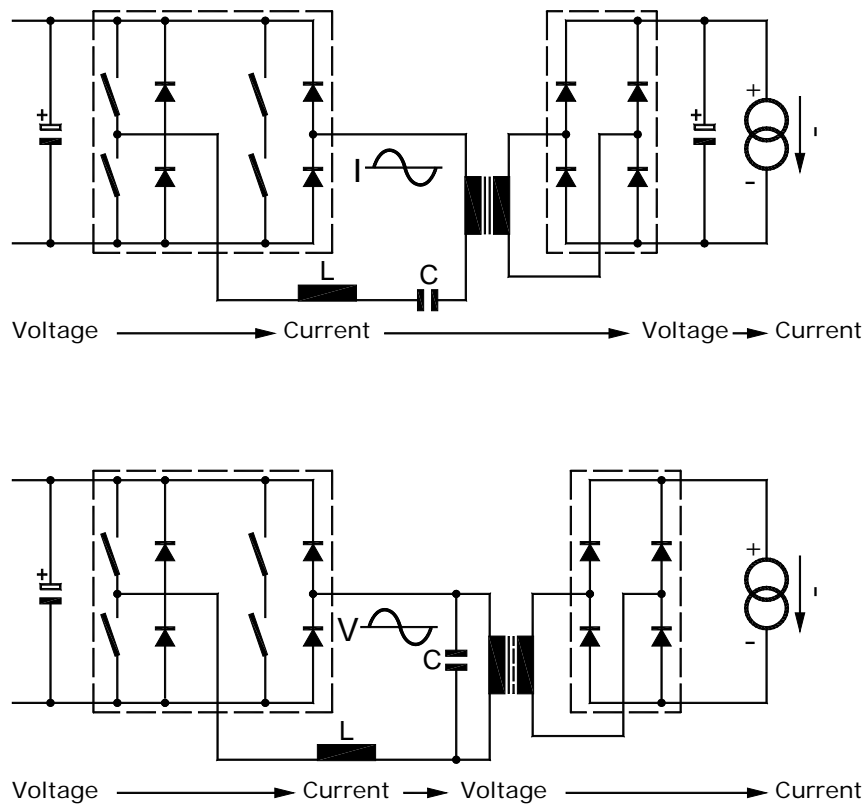

**Fig. 27:** Series (above) and parallel (below) resonant converters

# 7 Power converter classification

Figure 28 is a table summarizing the power conversion topologies. The crossed cells correspond to reversibility incompatibilities between the input and output sources. Two symmetric cells, with respect to the diagonal line D, represent two reversible topologies. Two topologies corresponding to two cells symmetric with respect to the point O are dual.

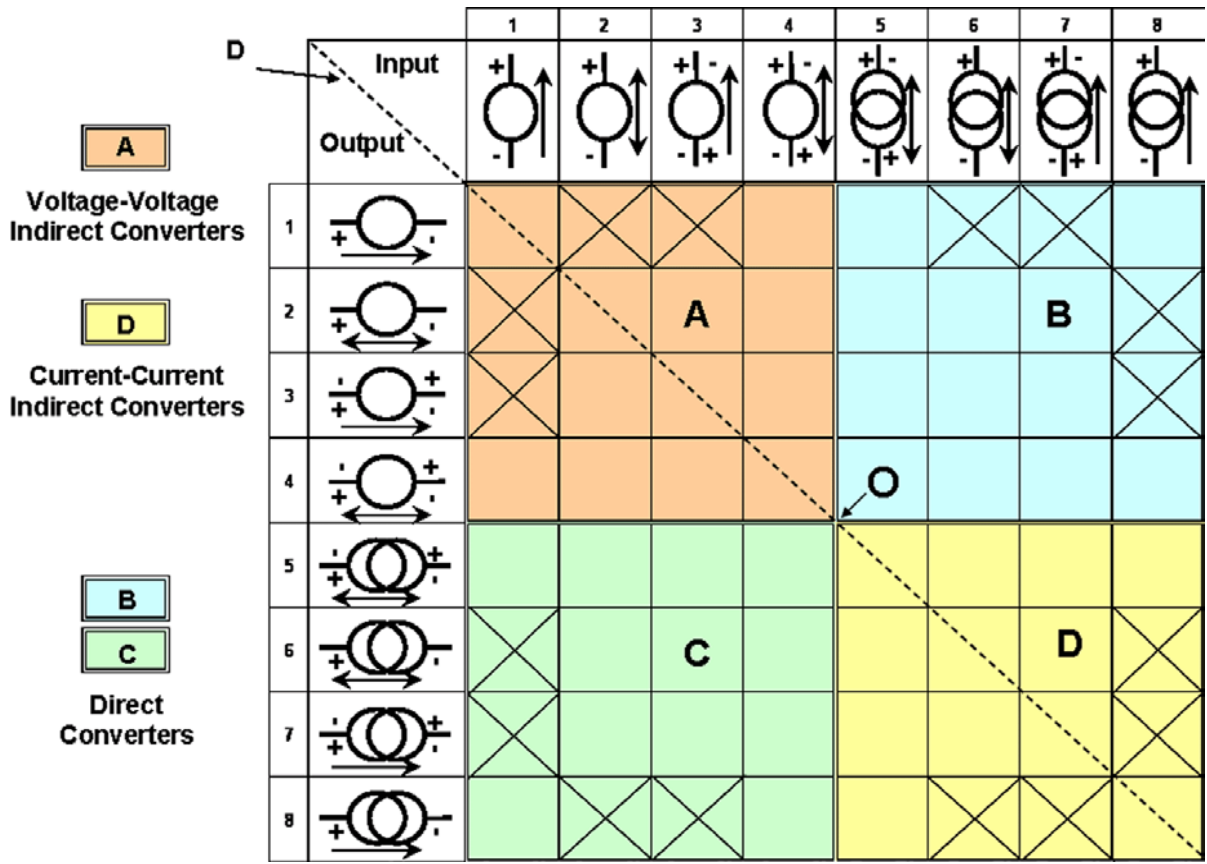

**Fig. 28:** General power conversion table

Figure 29 gives the different types of power converters and their usual names.

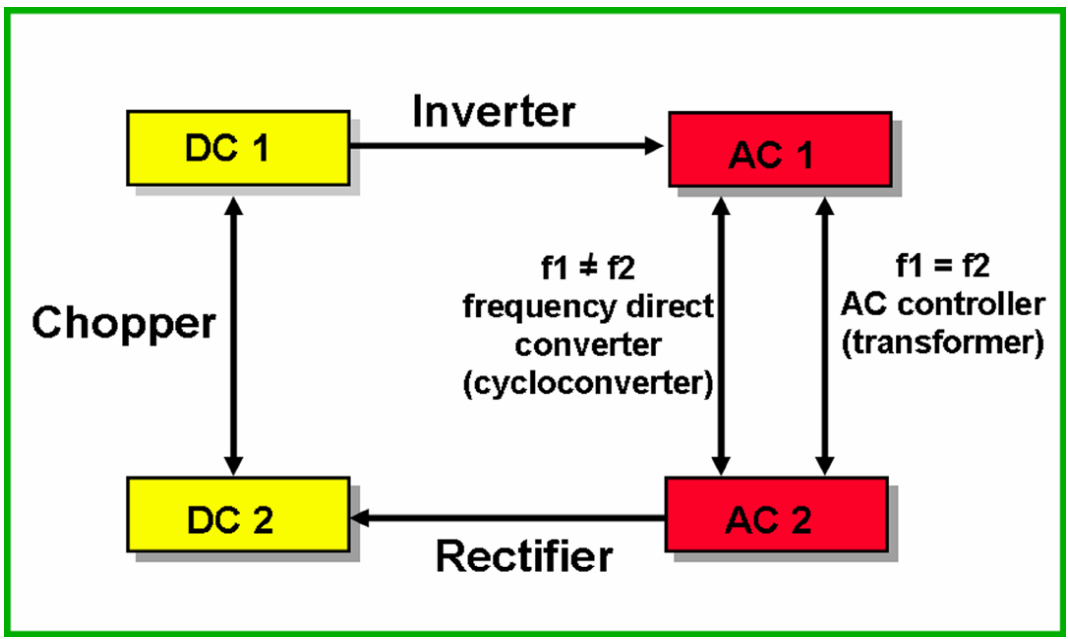

**Fig. 29:** Power converter classification

# 8 Synthesis of power converters

## 8.1 Synthesis method

A general and systematic method to obtain the power converter topologies is given below.

1. Determine the natures of the input and output sources. The basic structure can then be chosen (Fig. 24).
2. From the specification, deduce the voltage and current reversibilities of the input and output sources (one configuration from the general table (Fig. 28)).
3. From the basic structure, identify the different phases of operation according to the reversibilities and the energy transfer control. If necessary, simplify the configuration (short-circuit or open switches).
4. For the various phases, check the sign of the current through the ON switches and the sign of the voltage across the OFF switches: the static characteristics $I_k(V_k)$ of each switches are defined.
5. From the specification (desired output current and/or voltage functions), deduce the sequence of the different phases. For every commutation, represent the working point of each switch before and after the commutation.
6. From the static and dynamic characteristics, choose the type of switches (semiconductor type).

## 8.2 Examples

### 8.2.1 Non-reversible current chopper

Hypothesis:

- power conversion between a voltage source and a current source;
- these two sources are unidirectional in voltage and in current.

    Following the steps of the synthesis method:

1. a direct converter topology can be used;
2. cell (1,8) of the general table according to the source reversibilities;
3. Sequence 1 (active phase) and 2 (free wheel phase) from Fig. 30;
4. the different plots for the four switches for the two sequences are represented in Fig. 30;
5. from the previous plots, it can be easily deduced that:
    - K1 is a controlled two-segment switch (T switch) (Fig. 13);
    - K2 is an inverse D switch;
    - K3 is a short-circuit;
    - K4 is an open-circuit.

The topology of the converter is thus fully determined.

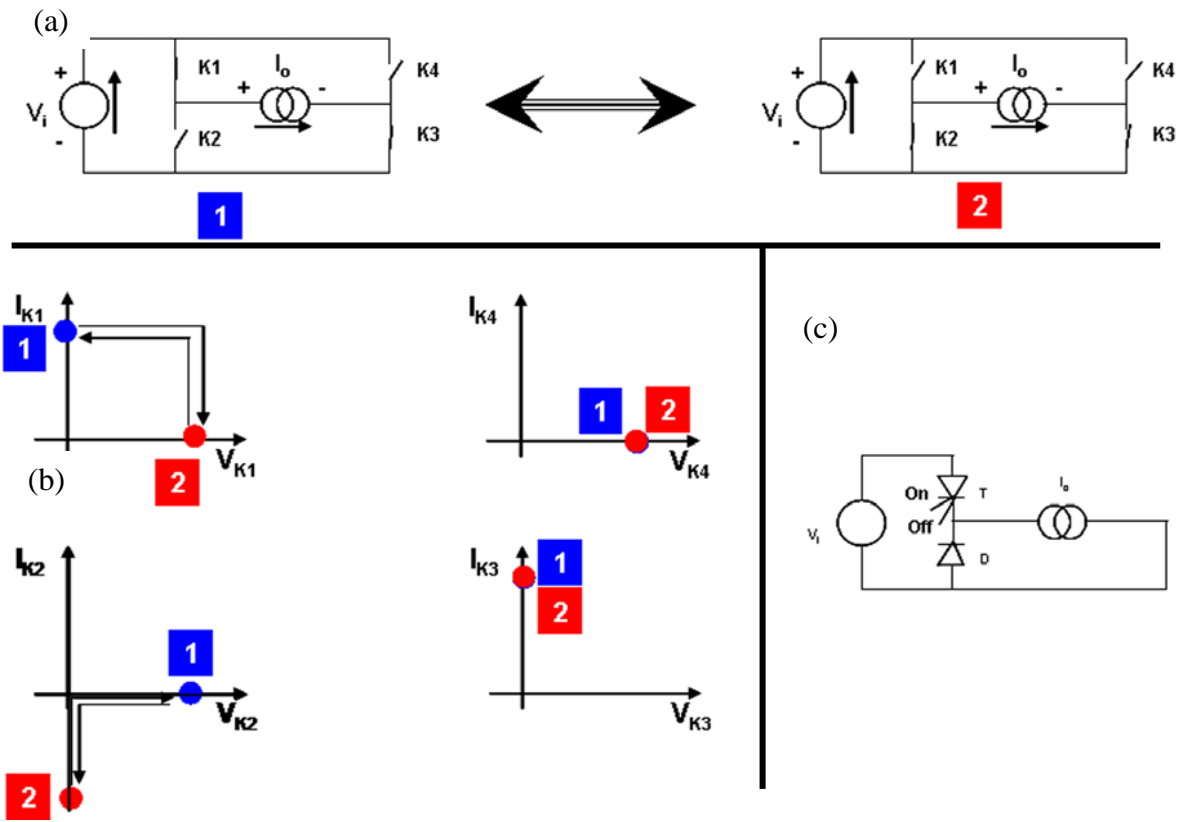

**Fig. 30:** Study of a non-reversible current chopper. (a) – sequences 1 and 2 (change in K1, K2, and K4); (b) – Current-voltage characteristics of each switch for the two sequences; (c) corresponding topology with transistor and diode.

*8.2.2 Reversible current chopper*

Hypothesis:

- power conversion between a voltage source and a current source;
- the input voltage source is bidirectional in current;
- the output current source is unidirectional in voltage and bidirectional in current.

  Following the steps of the synthesis method:

1. a direct converter topology can be used;
2. cell (2,6) of the general table according to the source reversibilities;
3. sequence 1 (active phase) and 2 (free wheel phase) of Figs. 30 and 31;
4. the different plots for the four switches are represented in Fig. 30 for the two sequences for the transfer of energy from the input source to the output source. Figure 31 represents the analysis of the brake phase: transfer of energy from the output source to the input source.
5. the analysis of the two phases leads to the determination of the structure presented in Fig. 32.

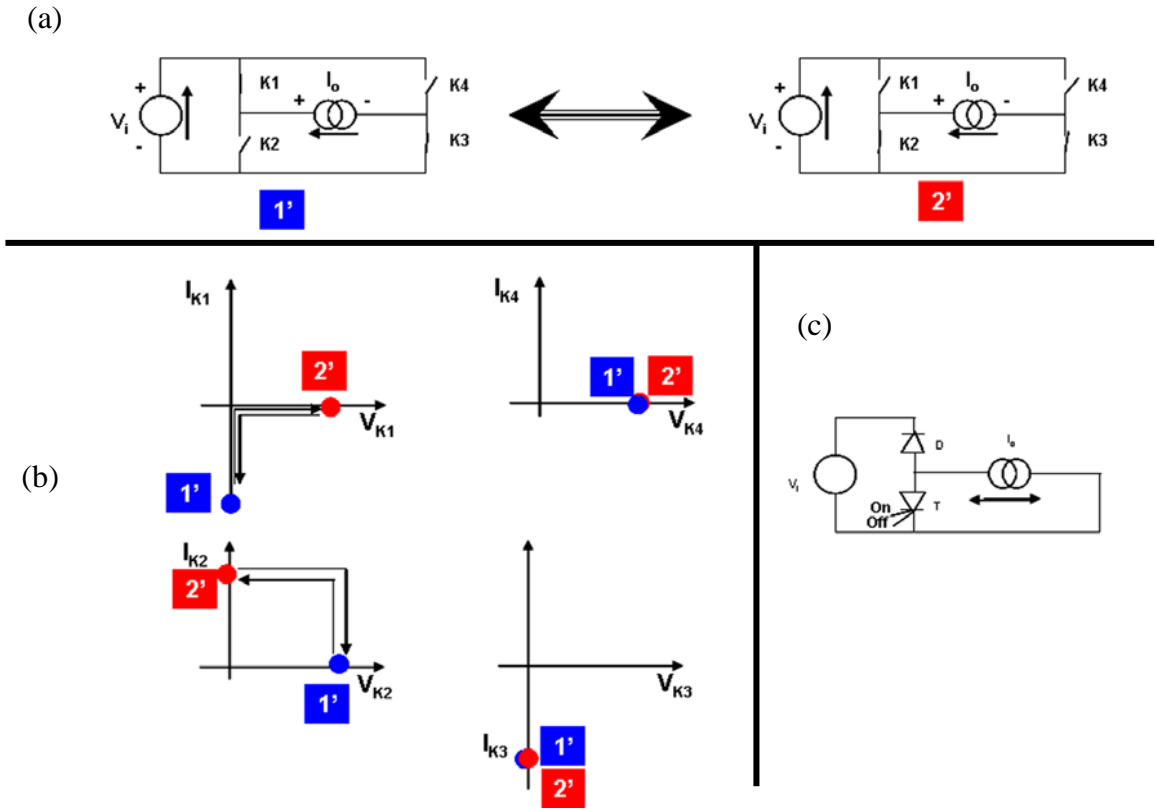

**Fig. 31:** Study of a reversible current chopper: brake phase. (a) – sequences 1 and 2 (change in K1, K2, and K4); (b) – Current-voltage characteristics of each switch for the two sequences; (c) corresponding topology with transistor and diode.

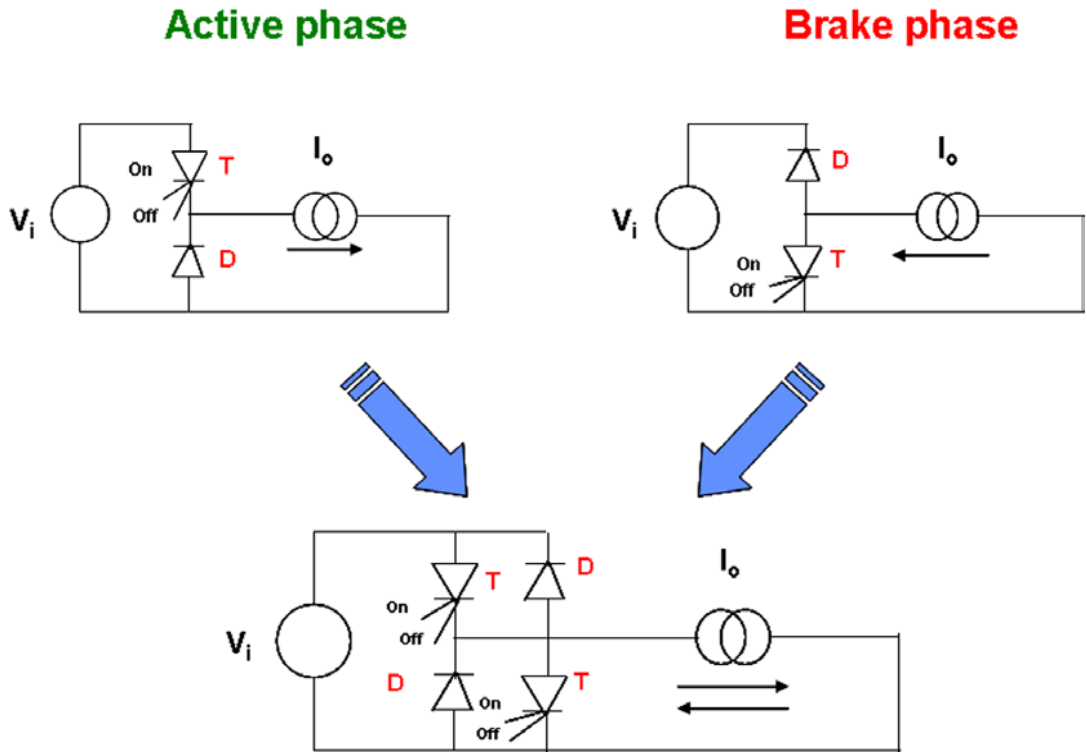

**Fig. 32:** Structure of reversible current chopper

*8.2.3 Voltage inverter*

Hypothesis:

- power conversion between a voltage source and a current source;
- the input voltage source is unidirectional in voltage and bidirectional in current; it is a DC source of value *E*.
- the output current source is bidirectional in voltage and bidirectional in current; it is an AC source. The specification is to get a +*E*, −*E* voltage at the output of the converter.

  The steps of the synthesis method are followed.

1. A direct converter topology can be used.
2. Cell (2,5) of the general table according to the source reversibilities.
3. The two sequences are shown in Fig. 33.
4. From the analysis of two sequences, it can be deduced that the switches must be bidirectional in current and unidirectional in voltage. They correspond to three-segment switches (D and T in parallel; Fig. 10).
5. To represent the working point of each switch, it is necessary to detail the specification. Two cases are defined according to whether the output voltage is ahead or not of the output current.

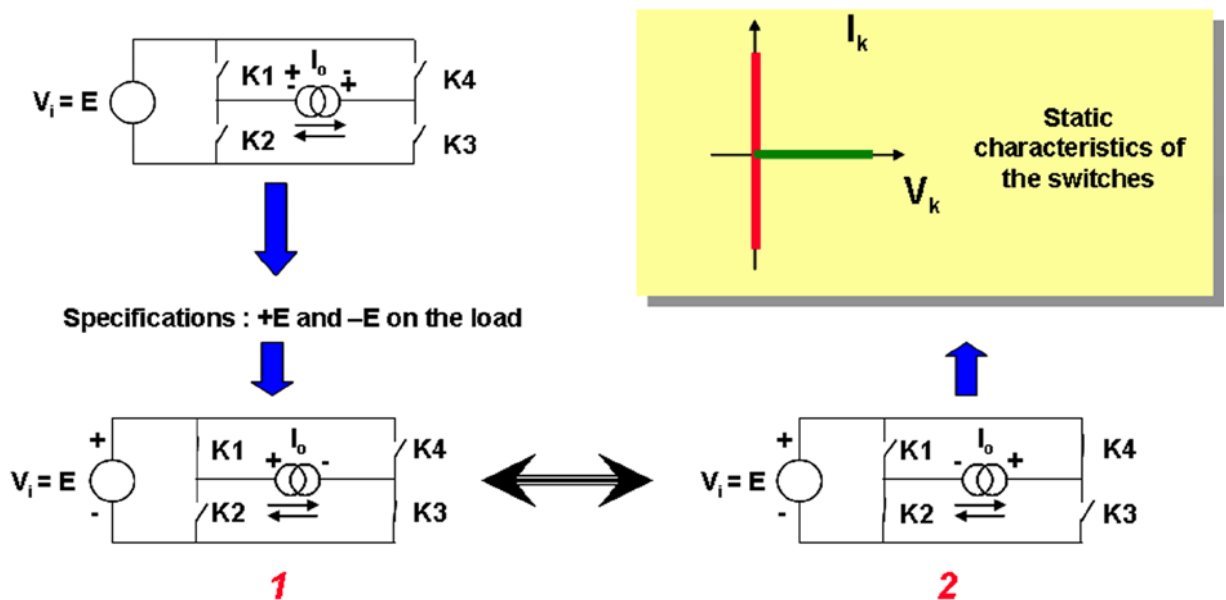

**Fig. 33:** Sequence of a voltage inverter

*8.2.3 .1  Case 1: The output voltage is ahead of the output current*

For the hypothesis that the output voltage is ahead of the output current, Fig. 34 shows the dynamic characteristics based on the analysis of the commutation between the two sequences.

It can be easily deduced that the four switches (K1, K2, K3, and K4) should be controlled turn-OFF and spontaneous turn-ON switches (dual thyristor; Fig. 15).

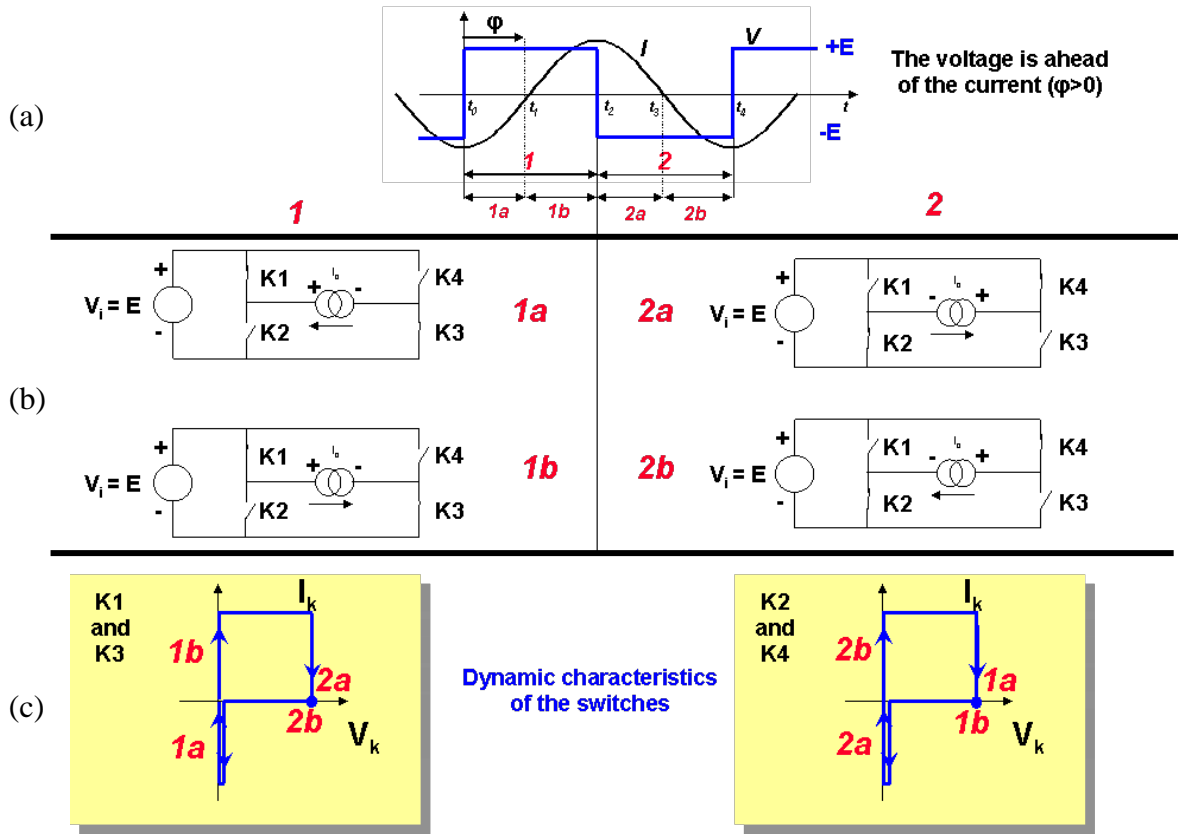

**Fig. 34:** Voltage inverter (Case 1): (a) – output voltage and current; (b) – Switches states for each time period (1a, 1b, 2a, and 2b); (c) - dynamic characteristics of the switches.

The topology of the converter is fully determined and is shown in Fig. 35. It is a zero-voltage-switching topology (see Section 9).

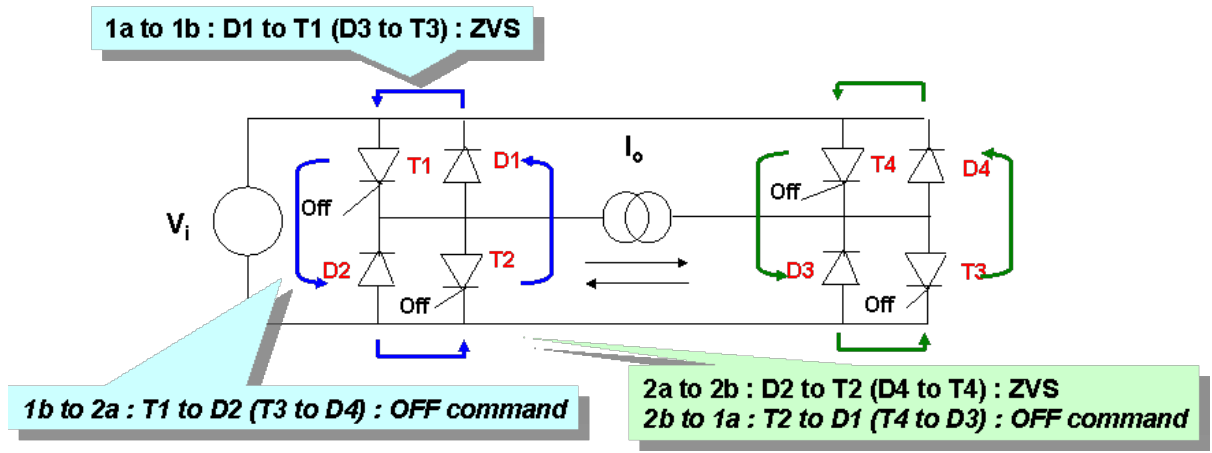

**Fig. 35:** Voltage inverter (Case 1): converter topology

*8.2.3 .2  Case 2: The output current is ahead of the output voltage*

Identical study can be made if the output current is ahead of the output voltage. Figures 36 and 37 represent the dynamics characteristics of the switches and the deduced topology. It is a zero-current-switching topology using a thyristor with a reverse diode in parallel (Fig. 15).

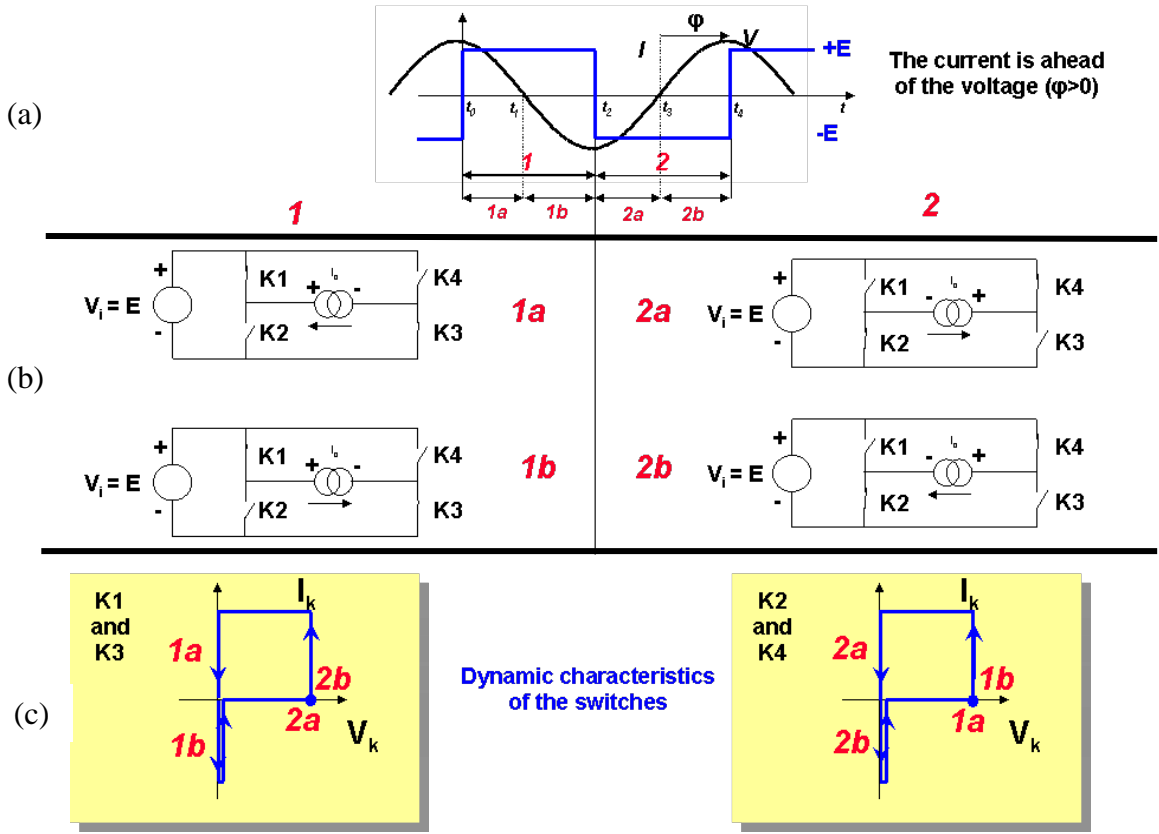

**Fig. 36:** Voltage inverter (Case 2): (a) – output voltage and current; (b) – Switches states for each time period (1a, 1b, 2a, and 2b); (c) - dynamic characteristics of the switches.

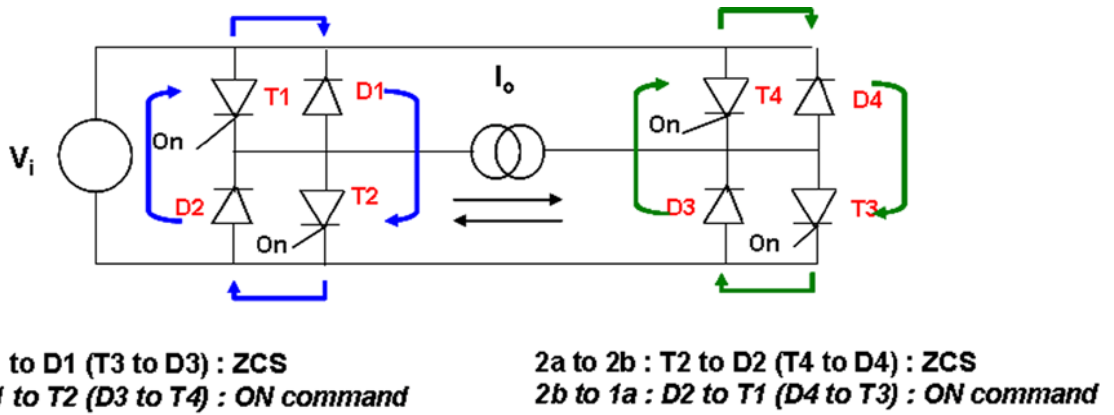

1a to 1b : T1 to D1 (T3 to D3) : ZCS
1b to 2a : D1 to T2 (D3 to T4) : ON command

2a to 2b : T2 to D2 (T4 to D4) : ZCS
2b to 1a : D2 to T1 (D4 to T3) : ON command

**Fig. 37:** Voltage inverter (Case 2): converter topology

## 9   Commutation cell

As presented in the previous chapters, the operation of a static converter can be split into sequences. A distinctive electrical network characterizes each sequence. The interconnection modifications of the sources by switches give the electrical networks. In the general case, the modifications are made by switches, which connect *n* branches to *one*. Figure 38 shows an example of a three-way switch.

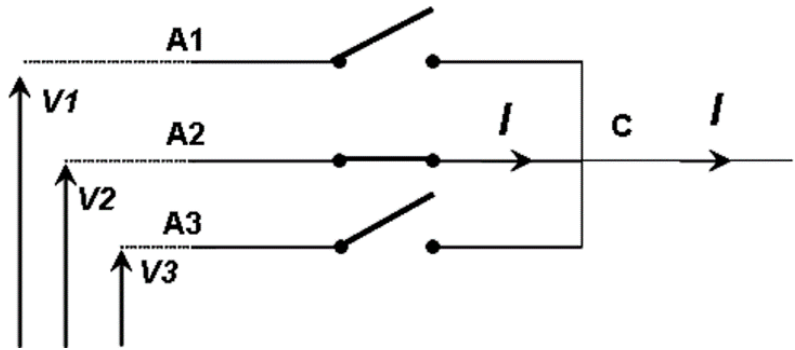

**Fig. 38:** Three-way switch

The network branches connected to these switches must fulfil the connection laws of the sources. Therefore, it can be deduced that:

– each switch is connected to a voltage source (otherwise opening a switch would result in open-circuiting a current source);

– the node at the centre of the star is connected to a current source since a voltage source can be connected only to a current source through a controlled switch;

– at a given time one and only one switch must be ON to avoid connecting two voltages sources and open-circuiting the current source.

Following these deductions, each commutation mechanism is a sequence of commutation only involving two switches. Thus, an elementary commutation cell, represented in Fig. 39, could be defined. The reversibilities of the voltage and current sources determine the static characteristics of the two switches. The switches need to have static characteristics with the same number of current and voltage segments. The two switches must be complementary, that is to say if one switch is ON the other one is OFF and, furthermore, if the turn-ON (turn-OFF) commutation of a switch is controlled, the turn-OFF (turn-ON) commutation of the other must be spontaneous.

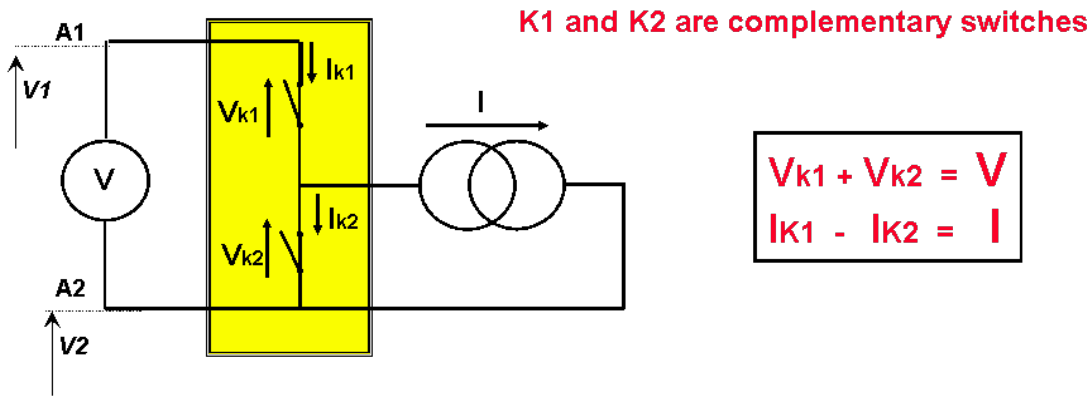

**Fig. 39:** Elementary commutation cell

To study a power converter topology, it is always fundamental to isolate all the elementary commutation cells and to check the complementarities of the switches. Figure 40 represents several examples of power converter topologies for which the elementary cells were highlighted.

Detailed information on commutation cells and on local and system commutation mechanism can be found in Refs. [9, 10, 13].

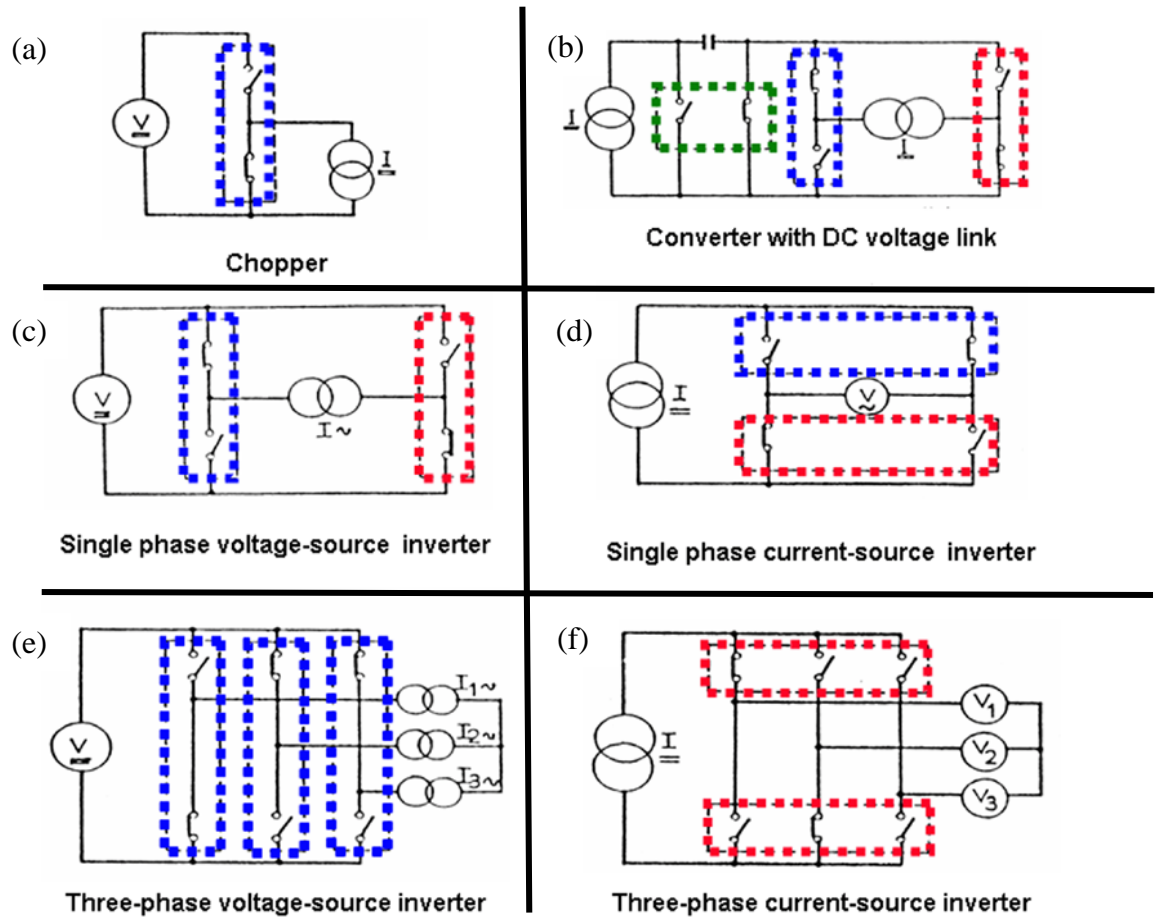

**Fig. 40:** Elementary cells in power converter examples. (a) – Chopper; (b) – indirect converter with DC voltage link; (c) – single phase voltage source inverter; (d) – single phase current-source inverter; (e) – three-phase voltage-source inverter; (f) – three phase current-source inverter.

## 10   Hard and soft commutation

In the domain of power conversion, the only available active components were diodes and thyristors. The topologies had to respect the commutation of these components: i.e. spontaneous turn-OFF of thyristors. The commutation mechanism was called 'natural commutation'.

If the spontaneous turn-OFF was not directly fulfilled (especially in the case of DC sources), it was necessary to add auxiliary circuits with reactive components (inductors and capacitors) and auxiliary semiconductors. The goal of these circuits was to create the conditions for thyristor turn-OFF. This commutation mechanism was called 'forced commutation'.

Power specialists were dreaming of having a power semiconductor with a controlled turn-OFF to avoid adding complex and costly auxiliary circuits and being limited to the grid frequency. With the development of the power transistor and the GTO, a new spectrum of topologies was opened. Then, specialists pushed to get faster components with simple and lighter drivers. From 1985, these components are available on the market. Power converters with higher power and frequency were then designed and built. However, the consequence is the management of (very) high d$V$/d$t$ and d$I$/d$t$ with stress on semiconductors but also on all of the other components. EMC became a constant concern of power electronics designers. Hard commutation was born! A first approach was to add components (snubber) to slow down the commutation (series inductor for a controlled turn-ON and a parallel capacitor for a controlled turn-OFF) and to avoid as much as possible the commutation losses close to the axes: aided-commutation. To avoid discharging the stored energy of the snubber in the semiconductor at the next commutation (inductance energy at turn-OFF and capacitor energy at turn-ON), it is necessary to add an auxiliary circuit to discharge the snubber energy and then to generate losses.

The solution to avoid all these problems and be able to use these new fast semiconductors with lossless snubbers is to have only one controlled command per switch. One commutation is spontaneous and the other one is controlled. Two types of commutations are possible: zero-voltage switching (ZVS; Fig. 41) and zero-current switching (ZCS; Fig. 42).

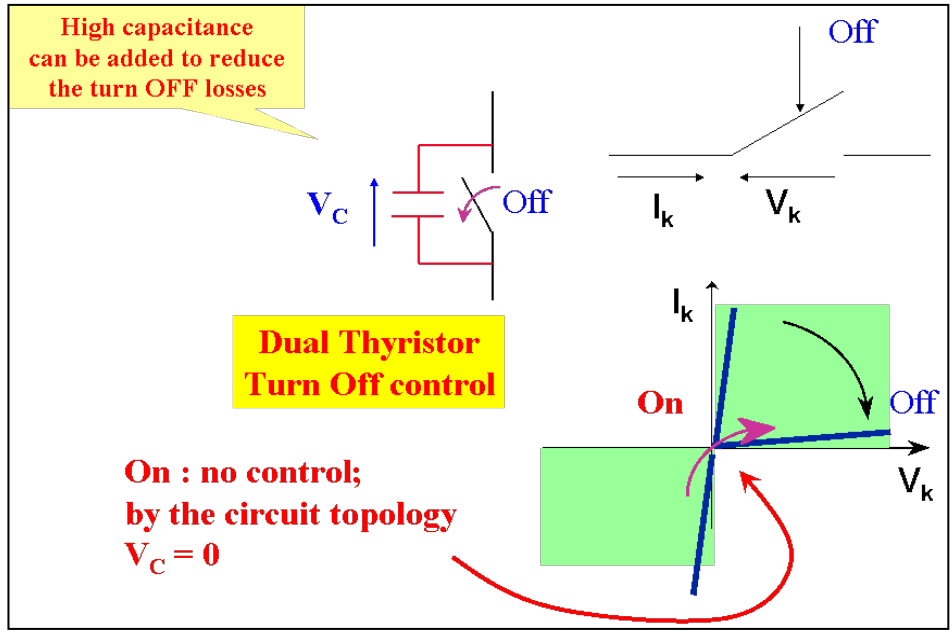

**Fig. 41:** Zero-voltage switching principle

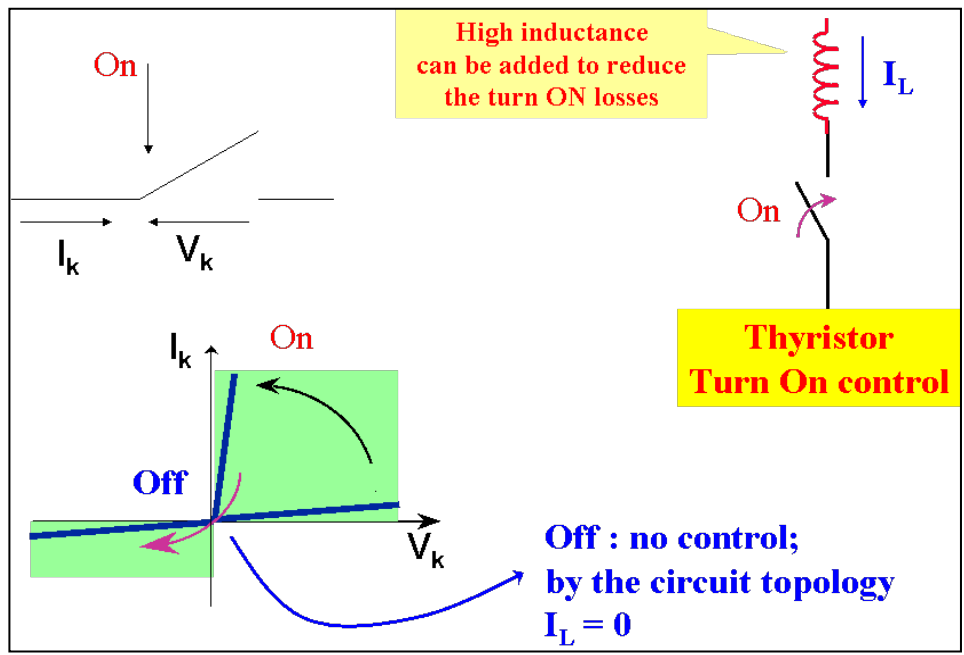

**Fig. 42:** Zero-current switching principle

The spontaneous commutation is lossless and it is easy to limit the losses for the controlled commutation with a series inductor for ZCS and a parallel capacitor for ZVS. These reactive components are naturally discharged for the next commutation, which is a spontaneous commutation. This commutation mechanism is called soft commutation.

For a power converter topology to obtain the conditions of soft-commutation, the following conditions must be fulfilled:

– the switches must have three-segment characteristics;

– the characteristics must be entirely described at each period;

– the converter must include reversible source(s) able to provoke the conditions for spontaneous commutation of the switches at the right instant.

The attractive properties of soft-switching [9, 11] are:

– the large reduction of switching losses;

– the improved reliability due to reduced stress;

– a limited frequency spectrum, which means an advantage with respect to EMI and losses in passive components;

– a reduction of weight and volume of the components resulting from the higher switching frequency;

– a higher bandwidth resulting from the high internal switching frequency;

– integration of parasitic elements in the commutation mechanism (e.g. leakage inductance of the transformer in the resonant circuit).

The soft-commutation domain has been a key research domain in power electronics in the last two decades. Numerous papers and conferences can be consulted for more information.

It should be noted that all LHC power converters were designed with soft-commutation topologies [12, 13, 14].

## 11  Conclusions

This paper has made an introduction and classification of the basic power converter components: sources and switches. From interconnection rules of the sources, the direct and indirect power converter topologies were deduced. A general and systematic method to synthesize was described and illustrated with examples.

With the fast development of turn-off controllable power semiconductors, the commutation mechanism becomes more and more important. To improve the performance of converters, frequencies are increased with the minimization of losses and EMI perturbations. Local treatment of commutation is no longer possible and it is crucial to design a suitable topology for the commutation of high-frequency and high-power semiconductors. To reach these goals, it is necessary to create, through the circuit topology, the turn-ON and turn-OFF conditions for the switches. Soft commutation is certainly the most appropriate and optimal solution.

### Acknowledgements

For the work reported here, different references in the domain of power electronics and especially from publications of the Laboratoire d'Electrotechnique et d'Electronique Industrielle, Toulouse, France (LEEI) members were used. The first author would like to express his deep gratitude and admiration to Professor Foch, who initiated the work around the systematic synthesis of power converter topologies. It was an opportunity and honour to work with him for more than 10 years.